\documentclass[twocolumn,twocolappendix]{aastex631}

\usepackage{natbib}

\usepackage{lineno} 

\usepackage{graphicx}

\begin{document}

\thispagestyle{empty}
~

\title{The Mass Dependence of the Fundamental Metallicity Relation in Observations and Simulations}

\correspondingauthor{Laura Carnevale} \\
\email{lcarnev@bu.edu}

\author[0009-0001-2861-7542]{Laura Carnevale}
\affiliation{Department of Astronomy, Boston University, 725 Commonwealth Avenue, Boston, MA 02215, USA}
\affiliation{Department of Astronomy, University of Virginia, Charlottesville, VA 22904, USA}
\affiliation{Virginia Institute for Theoretical Astronomy, University of Virginia, Charlottesville, VA 22904, USA}

\author[0000-0002-8111-9884]{Alex M. Garcia}
\affiliation{Harvard-Smithsonian Center for Astrophysics, 60 Garden Street, Cambridge, MA 02138, USA}
\affiliation{Department of Astronomy, University of Virginia, Charlottesville, VA 22904, USA}
\affiliation{Virginia Institute for Theoretical Astronomy, University of Virginia, Charlottesville, VA 22904, USA}
\affiliation{The NSF-Simons AI Institute for Cosmic Origins, USA}

\author[0000-0002-5653-0786]{Paul Torrey}
\affiliation{Department of Astronomy, University of Virginia, Charlottesville, VA 22904, USA}
\affiliation{Virginia Institute for Theoretical Astronomy, University of Virginia, Charlottesville, VA 22904, USA}
\affiliation{The NSF-Simons AI Institute for Cosmic Origins, USA}

\author[0000-0002-1768-1899]{Sara L. Ellison}
\affiliation{Department of Physics \& Astronomy, University of Victoria, Finnerty Road, Victoria, British Columbia, V8P 1A1, Canada}

\author[0009-0009-1792-7199]{Shweta Jain}
\affiliation{Department of Physics and Astronomy, University of Kentucky, 505 Rose Street, Lexington, KY 40506, USA}

\author[0000-0002-1333-147X]{Peixin Zhu}
\affiliation{Harvard-Smithsonian Center for Astrophysics, 60 Garden Street, Cambridge, MA 02138, USA}

\author[0000-0002-3247-5321]{Kathryn~Grasha}
\altaffiliation{ARC DECRA Fellow}
\affiliation{Research School of Astronomy and Astrophysics, Australian National University, Canberra, ACT 2611, Australia} 

\author[0000-0003-4792-9119]{Ryan L. Sanders}
\affiliation{Department of Physics and Astronomy, University of Kentucky, 505 Rose Street, Lexington, KY 40506, USA}

\author[0000-0001-6950-1629]{Lars Hernquist}
\affiliation{Harvard-Smithsonian Center for Astrophysics, 60 Garden Street, Cambridge, MA 02138, USA}

\author[0000-0001-8152-3943]{Lisa J. Kewley}
\affiliation{Harvard-Smithsonian Center for Astrophysics, 60 Garden Street, Cambridge, MA 02138, USA}

\author[0000-0002-8798-3972]{Jillian M. Scudder}
\affiliation{Department of Physics \& Astronomy, Oberlin College, 110 North Professor St, Oberlin, OH 44074, USA}

\begin{abstract}
The metal content of galaxies provides direct insight into the underlying physical processes that drive galaxy evolution.
An example of this is the three-parameter relationship between stellar mass, gas-phase metallicity, and star formation rate, commonly referred to as the Fundamental Metallicity Relation (FMR).
Previous studies have suggested that the FMR is redshift-invariant (at $z \lesssim 4$) and fully accounts for the scatter in the mass–metallicity relation (MZR).
In this work, we test this `fundamental' relation in both cosmological simulations (EAGLE, SIMBA, Illustris, IllustrisTNG) and Sloan Digital Sky Survey (SDSS) observations.
We find that the canonical anti-correlation between metallicity and specific star formation rate (sSFR) inverts in massive galaxies~($M_\star\gtrsim10^{10.5}~\mathrm{M}_\odot$) in EAGLE, IllustrisTNG, and SDSS.
When including lower star forming galaxies, the positive correlation appears for all four simulations and SDSS.
We speculate that this inversion may being driven by strong nuclear outflows (from, e.g., active galactic nuclei or stellar feedback), which quench star formation while simultaneously expelling preferentially enriched gas from the center of the galaxy.
We also find that this `inversion' appears in a number of metallicity diagnostics in observations~(though the details depend on diagnostic) and persists out to $z \sim 1$ in the simulations.
These results demonstrate that these strong nuclear outflows challenge simple gas regulator-type models and provide a new framework to test models of the baryon cycle in both future simulations and observations.
\end{abstract}


\section{Introduction} \label{sec:intro}

The chemical enrichment of galaxies provides a direct record of their evolutionary histories. 
Metals are produced in stellar interiors during nuclear fusion and released into the interstellar medium (ISM) through stellar winds and supernova explosions \citep{Lacey_1985, Koeppen_1994, Friedli_1994}. 
Meanwhile, galaxies accrete metal-poor gas from the circumgalactic medium, fueling new star formation and diluting the ISM \citep{Keres_2005, Dekel_2006, McKee_2007, Kennicutt_2012}. 
This baryon cycle of gas inflows, outflows, star formation, and metal production links a galaxy’s metal content (metallicity) to the balance of these processes \citep[e.g.,][]{Davé_2011, DeRossi_2015, Torrey_2018, Kewley_2026}.

Perhaps the most clear example of this interplay in action is the stellar mass gas-phase metallicity relation (MZR).
The MZR shows that galaxies with larger stellar mass generally correlate with higher gas-phase metallicity \citep{Lequeux_1979, Tremonti_2004}, a trend observed out to high redshifts \citep{Savaglio_2005, Zahid_2014, Sanders_2018, Sarkar_2025, Jain_2025}. 
Within the scatter of the MZR exists a tertiary correlation with star formation rate (SFR) such that galaxies with lower metallicities tend to have higher SFRs (and vice versa).
The relation is sometimes referred to as the Fundamental Metallicity Relation (FMR; although see discussion below; \citeauthor{Ellison_2008} \citeyear{Ellison_2008}, \citeauthor{LaraLopez_2010} \citeyear{LaraLopez_2010}, \citeauthor{Mannucci_2010} \citeyear{Mannucci_2010}).
While the existence of an anti-correlation between metallicity and SFR (or gas content) has been confirmed across multiple surveys \citep[e.g.,][]{Bothwell_2013,Bothwell_2016,Zahid_2014, Sanders_2018, Nakajima_2023, Li_2023, Stephenson_2024, Curti_2024,Khostovan_2025}, open questions remain about possible dependencies on redshift \citep{Nakajima_2023, Garcia_2024, Garcia_2025, Curti_2024, Pistis_2024,Stanton_2026} or stellar mass \citep{Yates_2012}, as well as whether stellar mass and/or SFR are truly the ``fundamental'' drivers of metallicity \citep{Baker_2023, Ma_2024, Sanchez_Menguiano_2024, Bassini_2024,McClymont_2026}.

Numerical simulations provide an essential tool for disentangling the physical drivers of galaxy evolution.  
Large-volume cosmological hydrodynamic simulations can sample tens of thousands of galaxies across a broad mass range while self-consistently modeling gas flows, star formation, and feedback.  
The MZR, and its anti-correlation with SFR/gas content, have been reproduced in such simulations \citep[e.g.,][]{Davé_2011, DeRossi_2015, Segers_2016, Torrey_2018, Torrey_2019, Garcia_2024, Garcia_2025}, enabling controlled tests of the impact of feedback physics. 
Recent work has shown evidence for FMR redshift evolution in several of these models \citep{Garcia_2024, Garcia_2025} and suggested a possible mass dependence in EAGLE \citep{DeRossi_2017}, potentially linked to feedback from active galactic nuclei (AGN; \citeauthor{Segers_2016} \citeyear{Segers_2016}).
If the FMR is truly a redshift-invariant and/or mass-invariant relation, current simulations may struggle to satisfy this tight constraint.

Moreover, there has yet to be a systematic quantification across simulation models, as well as with observations, of the mass dependence of the FMR.
It therefore remains unclear whether a mass-dependent FMR is a robust physical feature, how sensitive it is to galaxy selection methods, and what mechanisms might drive it.  
In this paper, we build on the work of \citeauthor{Garcia_2024} (\citeyear{Garcia_2024}) and \citeauthor{DeRossi_2017} (\citeyear{DeRossi_2017}) to address these questions by examining the behavior of the stellar mass--gas-phase metallicity--SFR relation in the EAGLE, SIMBA, Illustris, and IllustrisTNG cosmological simulations, as well as in Sloan Digital Sky Survey (SDSS) observational data. 
For coherence and clarity, in the rest of this work we distinguish between the {FMR} as it is generally referred to and a \textit{residual correlation (with SFR) about the MZR}.
Specifically, we use FMR to refer to the \textit{invariant} relation between stellar mass, gas-phase metallicity, and star formation rate.
When discussing specifically the relation between those same parameters that varies with mass or redshift, we refer to a residual correlation about the MZR.

The structure of our paper is as follows.
In Section \ref{sec:methods}, we describe the four different simulations and the SDSS data we analyze. 
We also explain the selection criteria for the galaxies used in our analysis. In Section \ref{sec:results}, we go over our results, showing that there is an inversion of the behavior of the gas phase metallicity--specific star formation relation as galaxies exceed a stellar mass of $10^{10.5} \mathrm{M}_\odot$. In Section \ref{sec:discussion}, we consider the effects of changing the galaxy selection criteria (\S \ref{eta2}) and the impact of the choice of metallicity diagnostic in SDSS (\S \ref{Z_diagnostic}).
We also expand our results to redshifts $0 < z < 1$ (\S \ref{redshifts}).
Lastly, we suggest that this inversion is caused by AGN feedback in high-mass galaxies (\S \ref{physics}). 
Finally, we summarize our findings in Section \ref{sec:conc}.

\section{Methods} \label{sec:methods}

In our analyses, we use data from four different cosmological simulations (Illustris, IllustrisTNG, EAGLE, and SIMBA) as well as observational data from SDSS. 
The nature of these simulations provides access to a large number of galaxies ($\sim10^5$) across a wide mass range ($M_\star \sim 10^8 - 10^{12.5}  \mathrm{M}_\odot$) which facilitates our study of the FMR.
In addition, we use four models that are similar in broader aspects (e.g., box size, numerical resolution) but different in others (detailed implementation of physical processes).
This enables meaningful comparisons of the resulting relations while also increasing confidence that any result common to all models reflects a genuine physical trend rather than being an artifact of the specific model implementation \citep[see also recent work by][etc]{Garcia_2024,Garcia_2024b,Garcia_2025,Garcia_2025b,Garcia_2026,Wright_2024,Lagos_2025}.
In the same vein, the advantage of using SDSS data is that we have access to information from thousands of observed galaxies across the same wide mass range as the simulations, giving us very similar samples to work with.
Comparing to observational data gives us a reference point on which we can benchmark simulation predictions against observations.
In the following section, we describe each of the models and the SDSS dataset, along with the selection criteria used in our analysis.

\subsection{Simulations}

\subsubsection{Illustris} \label{subsec:Illustris}
The Illustris cosmological simulation suite~\citep{Vogelsberger_2014a, Vogelsberger_2014b, Genel_2014, Sijacki_2015} is executed with the moving mesh code \textsc{arepo}~\citep{Springel_2010, Pakmor_Bauer_&_Springel_2011} using the original Illustris galaxy formation model~\citep{Vogelsberger_2013, Torrey_2014}.
It employs WMAP cosmology \citep{WMAP}: $\Omega_\Lambda$ = 0.73, $\Omega_m$ = 0.27, $\Omega_b$ = 0.0456, $H_0$ = 70.4 km/s/Mpc, and $\sigma_8$ = 0.81.
The simulations are boxes of side length 75 ${\rm Mpc}\,h^{-1}$, within an initial baryon mass resolution of $1.26\times 10^6\, \mathrm{M}_{\odot}$.
Illustris uses the \cite{Springel_Hernquist_2003} equation of state to model the dense interstellar medium, which allows star formation above a certain gas density threshold $(n_{\rm H}>0.13~{\rm cm}^{-3})$ while pressurizing the ISM gas against artificial collapse.
Stellar feedback is implemented via kinetic galactic winds that are launched from the star forming gas at a rate proportional to the star formation rate.
The velocity of the winds is scaled with the dark matter velocity dispersion.

Nine elements are tracked during the simulation: 
H, He, C, N, O, Ne, Mg, Si, and Fe. 
Specifically, chemical enrichment from type Ia and II supernovae, as well as enrichment from asymptotic giant branch stars are included, with metal yields taken from \cite{Thielemann_2003}, \cite{Portinari_1998}, and \cite{Karakas_2010}, respectively.

Black holes of mass $10^5 \, \mathrm{M}_{\odot} h^{-1}$ are seeded if and when the galactic halo mass reaches $5 \times 10^{10}\, \mathrm{M}_{\odot} h^{-1}$.
The AGN accretion rate is capped at the Eddington limit and the associated feedback is implemented via both a low and a high accretion mode.
The low accretion mode, or `radio mode', injects thermal `bubbles' within a radius of 50 kpc of the black hole.
The high accretion mode, or `quasar mode', injects thermal energy into the surrounding region.
The transition between these two modes is fixed at 5 percent of the Eddington rate.

\subsubsection{IllustrisTNG (TNG)} \label{subsec:TNG}

IllustrisTNG is a successor of the Illustris cosmological simulation \citep[`TNG' standing for The Next Generation;][]{Marinacci_2018,Naiman_2018,Nelson_2018,Pillepich_2018b,Pillepich_2018,Pillepich_2019,Springel_2018, Nelson_2019a, Nelson_2019b}.
For ease of reference and to prevent confusion with the original Illustris simulations, we will henceforth refer to IllustrisTNG simply as TNG.
TNG's physical processes have been modified compared to the original Illustris, to obtain a better match to observational results (\citeauthor{Weinberger_2017} \citeyear{Weinberger_2017}, \citeauthor{Springel_2018} \citeyear{Springel_2018}).
For the purposes of this work, we only briefly list a few of the relevant differences betewen the updated TNG model and that of Illustris; however, we refer the reader to \cite{Pillepich_2018} for a more complete description of differences.

The TNG simulations were also carried using the moving Voronoi mesh code {\sc arepo} (\citeauthor{Springel_2010} \citeyear{Springel_2010}, \citeauthor{Pakmor_Bauer_&_Springel_2011} \citeyear{Pakmor_Bauer_&_Springel_2011}).
The cosmology ($\Omega_\Lambda$ = 0.692, $\Omega_m$ = 0.31, $\Omega_b$ = 0.0486, $H_0$ = 67.7 km/s/Mpc, $\sigma_8$ = 0.8159, $n_s$ = 0.97) is adapted from the Planck Collaboration XIII data \citep{Planck_2016}. 
The model calibration was done using a combination of the stellar mass function, black hole mass vs. stellar mass at $z = 0$, the cosmic SFR density function, and the hot gas fraction in galaxy clusters at $z = 0$.
The simulations are carried out in boxes of various side lengths. 
Here, we analyze data from TNG100-1, the highest resolution $75~{\rm Mpc}\,h^{-1}$ volume, with an initial baryonic resolution of $1.4\times 10^6 \,\mathrm{M}_{\odot}$, both comparable to the original Illustris simulation \citep{Pillepich_2018}.

TNG keeps track of how the stellar population evolves, including enrichment from type Ia and II supernovae, as well as from the asymptotic giant branch using metal yields tables from \cite{Nomoto_1997}, \cite{Portinari_1998}, \cite{Kobayashi_2006}, \cite{Karakas_2010}, and \cite{Fishlock_2014}. 
In doing so, the same 9 elements as in Illustris (H, He, C, N, O, Ne, Mg, Si, and Fe) are tracked.\ignorespaces
\footnote{\ignorespaces
TNG also has a tenth ``other metals'' field which acts as a proxy for the metal species not explicitly tracked.
}
Star formation occurs when a certain density threshold is reached.
TNG uses the same \citeauthor{Springel_Hernquist_2003} (\citeyear{Springel_Hernquist_2003}) equation of state as Illustris.

The stellar feedback has both kinetic and thermal components, with 90 percent being injected as momentum and the remaining 10 percent as an increase in temperature.
This injection occurs isotropically.
The ejected feedback is temporarily decoupled from the hydrodynamic code. 
The winds recouple after either 2.5 percent of the Hubble time has elapsed, or when the particle reaches a cell below a certain density, whichever happens first.
Different from the original Illustris model, stellar feedback--driven winds in TNG have an effective wind speed floor of $350~{\rm km\,s}^{-1}$, which has the effect of increasing the efficacy of stellar feedback in low mass and high redshift systems.

Similar to the base Illustris simulation, if the halo of a galaxy exceeds $5 \times 10^{10}\,\mathrm{M}_{\odot}h^{-1}$, a black hole of mass $8 \times 10^5\, \mathrm{M}_{\odot} h^{-1}$ is seeded. 
These black holes can become more massive either by merging with other black holes or by accreting mass.
While TNG also has a low and a high modes for black hole feedback \citep{Weinberger_2017,Weinberger_2018}, they are notably distinct from the original Illustris model.
The transitional threshold from one to the other varies and depends on the mass of the black hole itself.
Critically, this black hole mass threshold roughly corresponds to a transition in galaxies with stellar mass of $\sim10^{10.5}\mathrm{M}_\odot$ \citep{Weinberger_2017}.
The lower accretion-rate feedback consists of kinetic energy from winds being added to the environment.
The higher accretion-rate feedback consists of a deposition of thermal energy around the black hole.
In both modes, the energy is injected isotropically. 
The injecta are not decoupled from the hydrodynamic code.

\subsubsection{SIMBA} \label{subsec:SIMBA}
SIMBA \citep{Davé_2019} is the spiritual successor of the MUFASA \citep{Davé_Thompson_&_Hopkins_2016} cosmological galaxy formation simulations. 
It uses the meshless finite mass mode of the {\sc gizmo} hydrodynamics code (\citeauthor{Hopkins_2015} \citeyear{Hopkins_2015}; \citeyear{Hopkins_2017}). 
The following cosmological parameters are used: $\Omega_\Lambda = 0.7$, $\Omega_m = 0.3$, $\Omega_b = 0.048$, $H_0 = 68$ km/s/Mpc, $\sigma_8 = 0.82$, and $n_s = 0.97$ \citep{Planck_2016}.
The simulations are carried out in boxes of side length $100~{\rm Mpc}\,h^{-1}$, the largest of the four simulations employed; however, the baryon mass resolution of $1.8\times 10^7\,\mathrm{M}_\odot$ is coarser than that of the other three simulations by a factor of $\sim10$ \citep{Davé_2019}. 
We therefore make slightly different resolution cuts for our galaxies in SIMBA compared to the other three simulation models (which we discuss further in Section~\ref{selection}).

SIMBA accounts for eleven elements (H, He, C, N, O, Ne, Mg, Si, S, Ca, Fe) in stellar populations. 
Enrichment is calculated by taking into account stellar evolution \citep{Iwamoto_1999} and type Ia \citep{Nomoto_2006} and II supernovae \citep{Oppenheimer_2006}.
The star formation itself takes into account $H_2$ densities, as guided by the Kennicutt-Schmidt law \citep{Kennicutt_1998}, using \cite{Krumholz_&_Gnedin_2011} for its equation of state.
It is worth noting that SIMBA also includes a model to track the production, growth, and destruction of dust \citep{Dwek_1998}.
As such, some of the metals in SIMBA are locked in dust.

The stellar feedback in SIMBA is modeled using decoupled two-phase winds, with thermal and kinetic components.
30 percent of the injected wind particles are injected as `hot', their temperature depending on the supernova energy and the wind's kinetic energy.
The injected wind is temporarily decoupled from the hydrodynamics code until its velocity relative to its environment is less than half of that environment's sound speed (\citeauthor{Anglés-Alcázar_2017b} \citeyear{Anglés-Alcázar_2017b}, \citeauthor{Muratov_2015} \citeyear{Muratov_2015}). 

A black hole of mass $10^4 \, \mathrm{M}_\odot h^{-1}$ is seeded when the galaxy halo reaches a mass of $10^{9.5} \mathrm{M}_\odot$ \citep{Davé_2019}.
SIMBA contains two AGN feedback modes: a kinetic and thermal jet-mode and a kinetic radiative mode.
Both modes inject winds into the surrounding space, although the winds of the jet boost have a much larger velocity (the radiative winds can reach velocities up to $1500~{\rm km\,s}^{-1}$, while the jet winds can reach up to $8000~{\rm km\,s}^{-1}$).
The jet-mode is present in black holes if they maintain a minimum mass of $10^{7.5} \mathrm{M}_\odot$  and an Eddington ratio of under 0.2.
In both modes, the winds are ejected as bipolar to the plane of the black hole.
The ejected winds remain decoupled from the hydrodynamic code for $10^{-4} t_{\rm H}(z)$, traveling for up to 10 kpc.
For cold particles, the accretion is limited by torques up to three times the Eddington limit (\citeauthor{Hopkins_&_Quataert_2011} \citeyear{Hopkins_&_Quataert_2011}, \citeauthor{Anglés_Alcázar_2017a} \citeyear{Anglés_Alcázar_2017a}); for hot particles, the accretion follows Bondi-Hoyle up to the Eddington limit \citep{Bondi_1952}.

\subsubsection{EAGLE} \label{subsec:EAGLE}
The EAGLE simulations (\citeauthor{Schaye_2015} \citeyear{Schaye_2015}, \citeauthor{Crain_2015} \citeyear{Crain_2015}) are cosmological simulations using a heavily modified version of {\sc gadget-3} \citep{Springel_2005} called {\sc anarchy} \citep{Schaller_2015}, which is a version of the N-body Tree-Particle-Mesh smoothed particle hydrodynamics solver, allowing it to calculate both short- and long-range particle interactions.

The cosmology ($\Omega_\Lambda$ = 0.693, $\Omega_m$ = 0.307, $\Omega_b$ = 0.04825, $H_0$ = 67.77 km/s/Mpc, $\sigma_8$ = 0.8288, $n_s$ = 0.9611) has been adapted from the Planck Collaboration XVI \citep{Planck_2014}.
EAGLE is calibrated using observations, including that of the stellar mass to black hole mass relation and SFR observations per \cite{Kennicutt_1998}.
The simulations are carried out in boxes of side length $67.8~{\rm Mpc}\,h^{-1}$, with a baryonic mass resolution of $1.8\times 10^6\,\mathrm{M}_\odot$ \citep{Schaye_2015}.

EAGLE explicitly keeps track of eleven different elements: H, He, C, N, O, Ne, Mg, Si, S, Ca, and Fe \citep{Schaller_2015}.
Metal yields are taken from \citeauthor{Thielemann_2003} (\citeyear{Thielemann_2003}; asymptotic giant branch), \citeauthor{Portinari_1998} (\citeyear{Portinari_1998}; type Ia supernovae), and \citeauthor{Marigo_2001} (\citeyear{Marigo_2001}; type II supernovae).
Star formation occurs in EAGLE according to a density threshold with an additional metallicity dependence, given by
\begin{equation}
    n_{\rm H} \geq 10^{-1}\,\left(\frac{Z}{0.02}\right)^{-0.64}~{\rm cm^{-3}}.
\end{equation}
The stellar feedback in EAGLE is purely thermal:
the feedback heats the surrounding gas to $10^{7.5}~{\rm K}$, meant to mimic the effects of stellar winds, radiation, and type Ia and II supernovae \citep{Dalla_Vecchia_&_Schaye_2012}.
EAGLE uses \cite{Schaye_Dalla_Vecchia_2008} for its equation of state, which is similar in spirit to that of \cite{Springel_Hernquist_2003}.

A black hole of mass $10^5~\mathrm{M}_\odot\,h^{-1}$ is seeded in the center of a galaxy if the galaxy's halo exceeds a mass of $10^{10}~\mathrm{M}_\odot\,h^{-1}$.
These black holes can gain mass either by merging with other black holes or by accretion following Bondi-Hoyle (\citeauthor{Bondi_1952} \citeyear{Bondi_1952}, \citeauthor{Springel_Di_Matteo_Hernquist_2005} \citeyear{Springel_Di_Matteo_Hernquist_2005}, \citeauthor{Schaye_2015} \citeyear{Schaye_2015}).
Feedback from these black holes is simulated in the form of an isotropic, thermal injection of $10^{8.5}$ K,
chosen to counteract rapid cooling.
The feedback rate from the AGN is determined by the accretion rate of the black hole itself.
For a more detailed description of AGN feedback in EAGLE, see \citeauthor{McAlpine_2017} (\citeyear{McAlpine_2017}).

For both stellar and black hole feedback, the heated cells are not decoupled from the hydrodynamic code.
This has the effect that mass, metals, energy, and momentum are transmitted locally, to neighboring cells, unlike in the other simulations---with decoupled stellar feedback and (for the most part) black hole feedback---where a wind particle's mass, metals, and energy end up in more distant gas cells.

\subsubsection{Simulated Galaxy Selection Criteria} \label{selection}

The galaxies used in the analysis are selected depending on their $z=0$ stellar mass and star formation rate.
We place an effective resolution cut by requiring all galaxies in our sample have at least $\sim10^2$ star particles.
In Illustris, TNG, and EAGLE, this corresponds to galaxies with total stellar masses above roughly $10^{8}~\mathrm{M}_\odot$. 
In SIMBA, which has a baryonic mass resolution approximately an order of magnitude worse than the other simulations, we select galaxies with stellar masses above $10^9~\mathrm{M}_\odot$.
The small number of galaxies at higher masses leads to substantial noise in the definition of the MZR and FMR.

\begin{figure}[ht!]
    \centering
    \includegraphics[width=\linewidth]{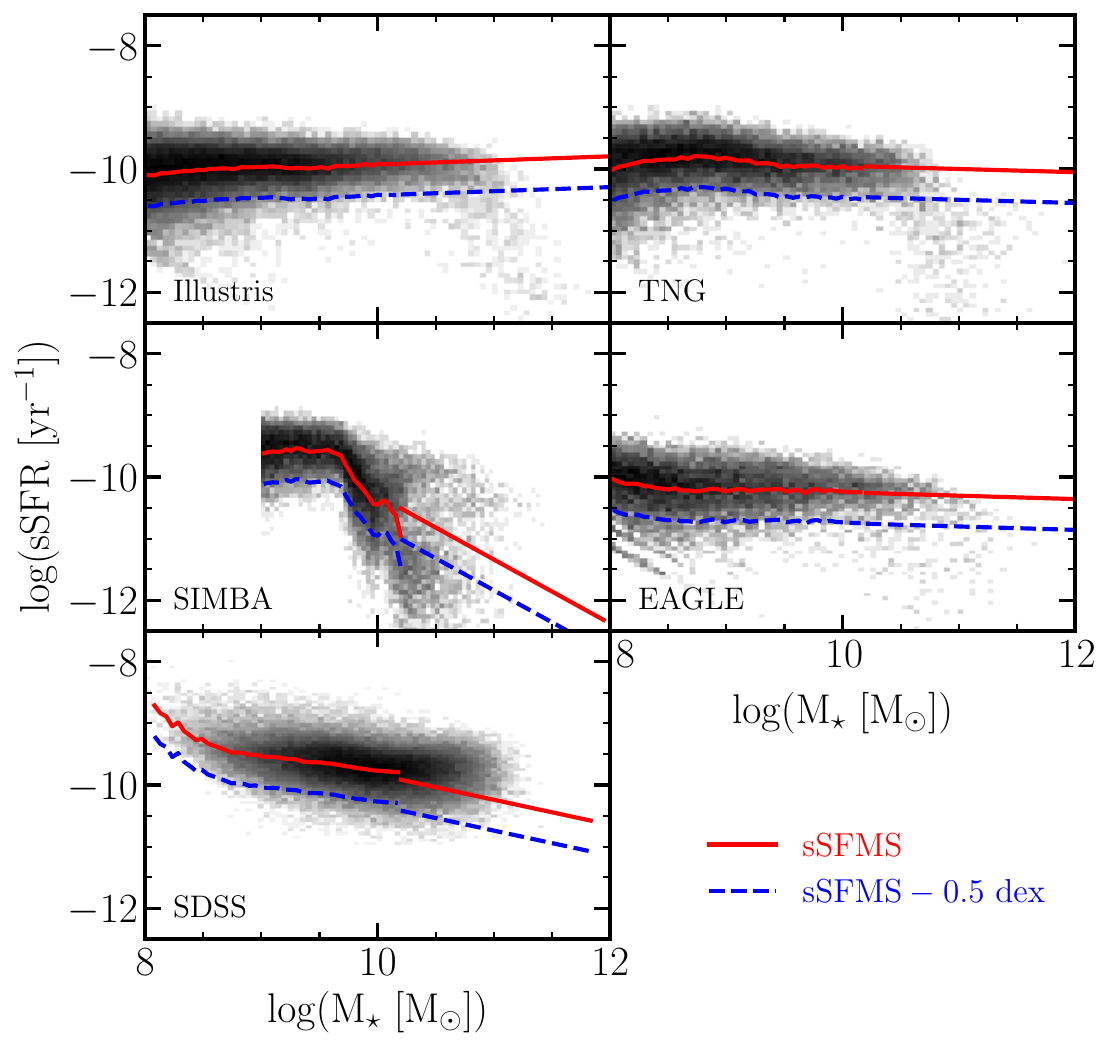}
    \caption{{\bf Galaxy Selection Criteria.}
    Specific star formation rate (sSFR) as a function of stellar mass for Illustris, TNG, SIMBA, and EAGLE simulations and SDSS observations.
    The median sSFR ($M < 10^{10.2} \mathrm{M}_\odot$) and expected sSFR at a given mass ($M > 10^{10.2} \mathrm{M}_\odot$) is plotted as a solid red line.
    The cut-off threshold of $0.5~{\rm dex}$ below that is shown as a dashed blue line.
    We investigate the impact of variations in our selection criteria in Section~\ref{eta2}.
    \label{fig:cut}
    }
\end{figure}

To obtain metallicity values from observations, strong emission lines from star forming regions are needed.
To evenhandedly compare simulations and observations, we place constraints on the SFR of the simulated galaxies.
The criteria selecting star-forming galaxies closely follow a number of previous simulation works \citep[e.g.,][]{Hemler_2021,Garcia_2023,Garcia_2024,Garcia_2025}.
We first select galaxies that have non-zero instantaneous SFRs.
We then define a specific star formation main sequence~(sSFMS) by binning the remaining galaxies in mass bins of $0.2~{\rm dex}$ and calculate the median specific star formation rate (sSFR; $\rm SFR/M_{\star}$) in each bin. 
For all of the galaxies below a stellar mass of $\rm 10^{10.2}~\mathrm{M}_\odot$, we define a threshold $-0.5$ dex below the median sSFR and select all galaxies above this threshold.
For galaxies with stellar masses above $\rm 10^{10.2} \mathrm{M}_\odot$, we perform a linear fit of the expected sSFR for the midpoint of the mass bin it belongs to and select those with sSFRs no more than $0.5$ dex below this line.
The fiducial cut-off value of $-0.5$ dex as well as the cut-off mass threshold of $10^{10.2} \mathrm{M}_\odot$ are adopted from \cite{Hemler_2021}. 
Figure \ref{fig:cut} shows the specific star formation rate as a function of stellar mass for the four simulations used in this analysis (as well as SDSS) with the median line plotted as a solid red line and the cut off line $0.5$ dex below it as a dashed blue line.
We discuss the impact of this fiducial choice in $0.5$ dex further in Section \ref{sec:discussion}.

\subsection{SDSS} \label{subsec:SDSS}

We compare our results from simulations against galaxy catalogs from the SDSS.
Specifically, the observational data for this analysis come from the seventh data release of the SDSS~\citep{SDSSDR7}.
The release includes spectral analyses of ~930,000 galaxies, as well as photometry in $u$, $g$, $r$, $i$, and $z$ bands.
The data were taken using a 2.5m telescope \citep{Gunn_2006} at the Apache Point Observatory in New Mexico.
We use stellar mass and SFR values from the MPA/JHU catalog \citep[described, e.g., in][]{Kauffmann2003, Brinchmann2004, Salim2007}.
Emission line fluxes were also taken from the MPA/JHU catalog, which are already corrected for Galactic extinction.  
We additionally apply a correction for internal extinction by assuming an SMC-like extinction law \citep{Pei_1992}.
For the bulk of this analysis, we use metallicity values (12 + log[O/H]) from \citeauthor{Curti_2017} (\citeyear{Curti_2017}; using their $\rm O_3N_2$ metallicity diagnostic).
In Section \ref{Z_diagnostic}, we compare different metallicity diagnostics and show that the choice of metallicity diagnostic has little qualitative impact on our results.
We select galaxies that have a signal-to-noise ratio of at least three in the four emission lines necessary for the BPT diagram \citep{Baldwin1981}.  

To match the samples from our simulation datasets, we select galaxies with total stellar masses between $10^{8}$ and $10^{12}~\mathrm{M}_\odot$.
Finally, we filter the observed galaxies in the same way as described for our simulated galaxies (see Section~\ref{selection}), defining a threshold $0.5$ dex below the sSFR main sequence (defined in the same fashion as in the simulations) and selecting all galaxies that fall above this threshold.\ignorespaces
\footnote{\ignorespaces
We note that this sSFMS cut on the observational sample is potentially redundant with the BPT selection.
However, we make this cut so as to make as fair a comparison as possible with the simulated datasets.
}

\section{Results} \label{sec:results}

\subsection{Comparison of MZRs} \label{fmr}

\begin{figure*}
    \centering
    \includegraphics[width=0.9\linewidth]{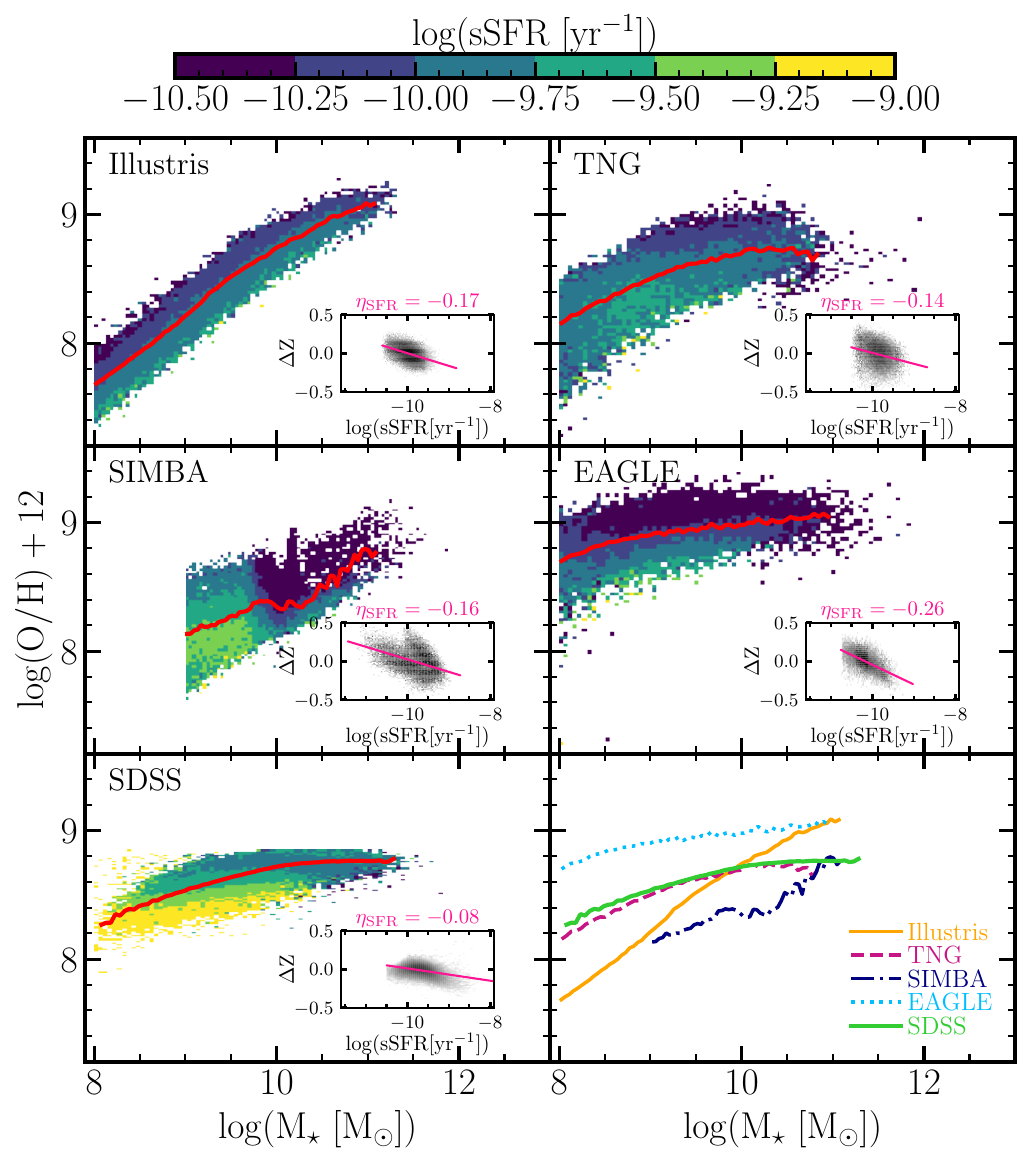}
    \caption{
    {\bf Comparison of MZR in Illustris, TNG, SIMBA, EAGLE, and SDSS.}
    Stellar mass of a galaxy versus gas-phase metallicity (mass metallicity relation; MZR) of that galaxy from Illustris, TNG, SIMBA, EAGLE, and SDSS. 
    The color gradient in each 2D histogram shows the distribution of specific star formation rate. 
    The solid red line shows the median MZR.
    The insets show galaxies' offset from the MZR as a function of sSFR for Illustris, TNG, SIMBA, EAGLE, and SDSS data. 
    A line of best fit is added, with its slope above each inset~(quoted as $\eta_{\rm SFR}$; see Equation~\ref{eq1}).
    The bottom right panel depicts all five median metallicity lines for visual clarity.
    We note that the absolute normalization of the MZR is difficult to compare directly between the simulations and observations~(e.g.,~\protect\citeauthor{Kewley_2008}~\protect\citeyear{Kewley_2008} and Appendix~\ref{appendix:diagnostics}, Figure~\ref{fig:diagnostics_MZR}).
    We therefore caution against a strict interpretation of the differences in the normalization of the MZRs.
    \label{fig:MZR}}
\end{figure*}

Figure \ref{fig:MZR} depicts the gas-phase metallicity as a function of stellar mass in each of the four simulations (Illustris, TNG, SIMBA, EAGLE) at $z=0$, as well as in SDSS data.
Bins in each 2D histogram are colored according to the average sSFR of galaxies within the bins.
In all five data sources, we find that there (i) is generally an increase in metallicity with increases in stellar mass and (ii) exists a correlation within the scatter with sSFR: at a given mass, galaxies with lower gas metallicities tend to have higher sSFRs~(and vice versa).

\subsubsection{Shape and Normalization}

We first investigate the normalization and shape of the MZRs in the four simulations and observations.
In each of the first five panels of Figure \ref{fig:MZR}, we plot the median MZR with a solid red line.
We summarize this information in the bottom right panel, which overlays each median MZR line.
We note that comparisons of the detailed normalization of the MZR are complicated by different metallicity diagnostics often providing different values~\citep[e.g.,][see also discussion in Section~\ref{Z_diagnostic}]{Kewley_2008}.
Nevertheless, it is remarkable to note how well the normalization of the SDSS and TNG MZRs align, they agree within $\lesssim0.1~{\rm dex}$ at virtually all stellar masses, despite no effort to calibrate directly to this relation.
This level of agreement in the normalization also speaks to the agreement in the shape of the TNG and SDSS MZRs.
EAGLE is systematically shifted by $\sim+0.4~{\rm dex}$ compared to SDSS~(comparable to previous findings in EAGLE, see, e.g.,~\citeauthor{Trayford_2019}~\citeyear{Trayford_2019}).
Despite the $0.4~{\rm dex}$ offset, the shape of the MZR in EAGLE has a similar level of agreement with SDSS as that of TNG.
The shape of the MZR is not as well reproduced in Illustris, however.
The median Illustris MZR appears much steeper than the SDSS relation.
As such, at low masses ($M_\star \lesssim 10^{10}~{\rm M}_\odot$), metallicities in Illustris tend to be slightly lower than in SDSS, whereas at higher masses ($M_\star\gtrsim10^{10}~{\rm M}_\odot$) they become higher than their SDSS counterparts.
Finally, galaxies in SIMBA tend to be on average $\sim0.25~{\rm dex}$ more metal-poor than those of SDSS at all $M_{\rm star}\lesssim10^{10.5}~{\rm M}_\odot$, beyond which they seem to begin to agree with SDSS more closely.
At the lower masses the MZR shape seems to agree better with SDSS, but the high mass up-turn yields a steeper relation.
These results are in good agreement with those of~\cite{Jain_2025}, who use a variety of observational data at $z=0-3$ to show that the shape of the MZR in TNG best matches that of their observations.

We note that the abrupt edge on the upper metallicity limit in the SDSS panel (bottom left) are due to the sample reaching the maximum viable metallicity for the metallicity calibration.

\subsubsection{Residual Correlation with sSFR} 

We next quantify the three-parameter relation between stellar mass, gas-phase metallicity, and SFR using a simple linear regression.
We obtain the offset of each galaxy's metallicity from the median MZR (offsets depicted in the inset panels of Figure \ref{fig:MZR}) and fit a linear regression (in logarithmic sSFR) such that 
\begin{equation} \label{eq1}
    \Delta Z = \eta_{\rm SFR} \log({\rm sSFR}) + c
\end{equation}
where $c$ is constant and $\eta_{\rm SFR}$ is the slope.
The value of $\eta_{\rm SFR}$ gives us a way of quantifying the offsets from the MZR as a function of the sSFR of the galaxy and is related to the commonly used projection of minimum scatter ``$\alpha$'' value used (see Equation~3 of \citeauthor{Garcia_2024} \citeyear{Garcia_2024}).
In this section we calculate $\eta_{\rm SFR}$ for all galaxies in each sample to give a baseline, before we investigate the behavior at different masses.
We find that $\eta_{\rm SFR}$ is negative for all five of our data sources when considering all of the galaxies in our samples.
Specifically, we obtain $\eta_{\rm SFR}=-0.17$ for Illustris, $-0.14$ for TNG, $-0.16$ for SIMBA, $-0.26$ for EAGLE, and $-0.08$ for SDSS.
Interestingly, the simulations generally have a stronger correlation between the offsets from the MZR and sSFR than that of observations.
It is at this point worth appreciating that despite there existing a three-parameter relationship in each sample, it is clear~(even by visual inspection of Figure~\ref{fig:MZR}) that the relationships are different in detail.
This has been highlighted previously in the simulations literature~\citep{Garcia_2024,Garcia_2025}, but the specific advantage of this work is putting it into context with the observational results.
Moreover, the qualitative agreement and quantitative disagreement---even on the full-population level---underscores the utility of the metal content of galaxies as a sensitive tracer of feedback physics.

\subsection{Stellar Mass Dependence of Scatter} \label{eta1}

Next, we investigate the mass dependence of the residual correlation about the MZR. 
To achieve this, we sort galaxies into stellar mass bins of width $0.25~{\rm dex}$ spaced every $0.25~{\rm dex}$ in logarithmic stellar mass.
In each of the mass bins, we find the offset from the median gas-phase metallicity.
We then perform the same linear fit to the data~(given by Equation~\ref{eq1}) and obtain an $\eta_{\rm SFR}$ value for each mass bin.

\begin{figure*}[t!]
    \centering
    \includegraphics[width=0.9\linewidth]{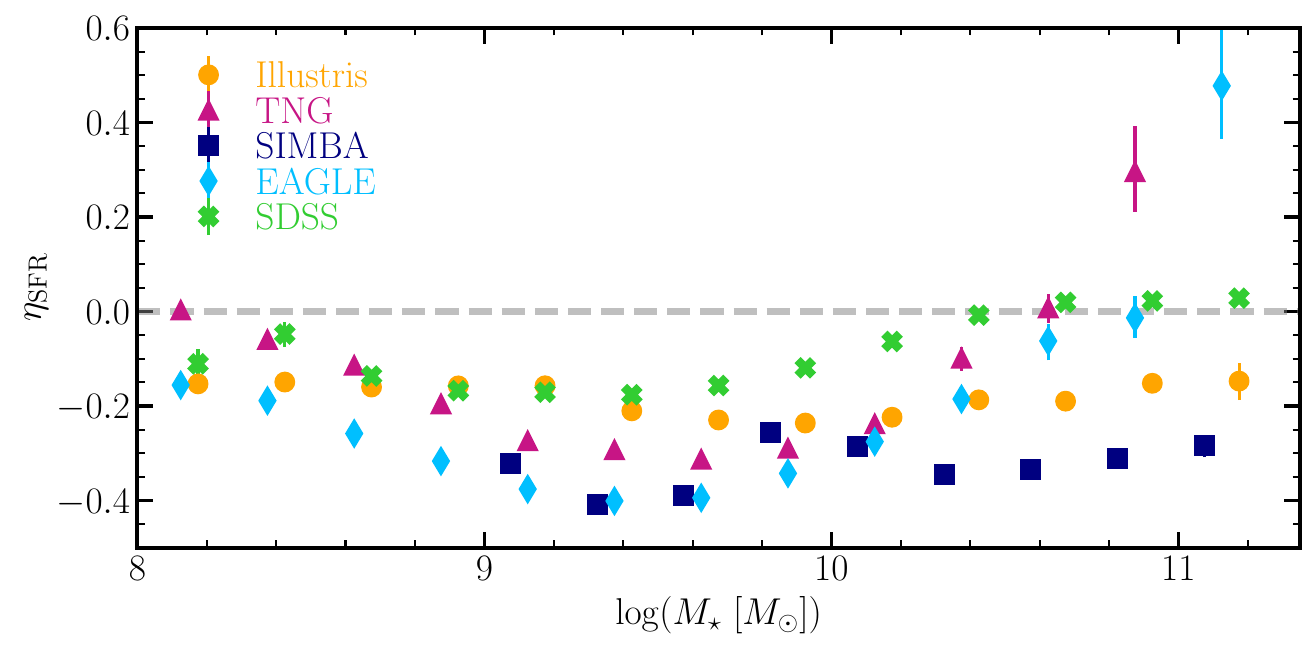}
    \caption{
    {\bf Correlation (or anti-correlation) of MZR offsets and sSFR as a function of stellar mass.}
    Comparison of the trend in offset from median metallicity as a function of sSFR ($\eta_{\rm SFR}$; see Equation~\ref{eq1}) for mass bins of width 0.25 dex for each of the four simulations used (TNG, Illustris, SIMBA, EAGLE) and SDSS.
    The dashed horizontal line indicates $\eta_{\rm SFR}=0$, where there is no correlation between offset from the MZR and sSFR.
    Point below this line show an anti-correlation between MZR offset and sSFR (e.g., the traditional FMR picture) and points above it show a direct correlation between MZR offset and sSFR.
    \label{fig:eta0.5}
    }
\end{figure*}

Figure \ref{fig:eta0.5} compares $\eta_{\rm SFR}$ values from the four simulations and SDSS as a function of mass.
We plot each $\eta_{\rm SFR}$ at the center of the mass bin~(with some light additional horizontal staggering for visual clarity).
We note that $\eta_{\rm SFR}$ values are only plotted if at least 50 galaxies are sampled in a particular mass bin.
As a measure of uncertainty on the derived $\eta_{\rm SFR}$ values we also perform a bootstrapping analysis.
Galaxies in each mass bin are randomly sampled (with replacement), we then calculate $\eta_{\rm SFR}$ within this newly generated random sample.
We perform this bootstrapping $1000$ times and quote the median, $16^{\rm th}$, and $84^{\rm th}$ percentiles of the distribution of $\eta_{\rm SFR}$ values in Figure~\ref{fig:eta0.5}.
Briefly, we find that, at lower masses ($M_\star\lesssim10^{10}~{\rm M}_\odot$), the uncertainties in $\eta_{\rm SFR}$ are negligible, as small changes in the selection of galaxies does not significantly change the overall distribution.
However, at higher masses ($M_\star\gtrsim10^{10}\mathrm{M}_\odot$), the smaller sample sizes mean that random sampling can have a larger effect on the sample, leading to more significant uncertainties.
Regardless, we note that our key results are not systematically impacted within the uncertainty of the bootstrapped samples.

For bins of stellar masses between $10^8$ and $\sim10^{10.5}~\mathrm{M}_{\odot}$, the $\eta_{\rm SFR}$ values are virtually all negative.
This suggests that lower sSFRs are correlated with higher metallicities~(and vice-versa), in good agreement with the results presented in the previous section.
We note that the lone exception at low mass is that of the $10^{8}-10^{8.25}\mathrm{M}_\odot$ stellar mass bin in TNG, which has an $\eta_{\rm SFR}$ consistent with $\sim0$.
This is likely due to the TNG model having a minimum wind floor, which serves to increase the efficiency of stellar feedback in low mass systems \citep{Pillepich_2018}.
We again emphasize that the lack of data points in SIMBA at masses below $10^9 \mathrm{M}_\odot$ is due to our galaxy selection criteria~(see Sections~\ref{subsec:SIMBA}~and~\ref{selection}).
Interestingly, there is some non-negligible mass dependence even within these mass bins that all have negative $\eta_{\rm SFR}$ values. 
The magnitude of $\eta_{\rm SFR}$ reaches its peak between $10^{9}~{\rm M}_\odot$ and $10^{10}~{\rm M}_\odot$ for each of Illustris, TNG, SIMBA, and EAGLE, whereas observational values are roughly consistent at all masses $(M_\star\lesssim10^{10}~{\rm M}_\odot)$.
It is worth appreciating that these systems are on the lower resolution end of these simulations.
It is therefore unclear whether the mass evolution in these smallest mass bins represents a physical disruption of the baryon cycle leading to an sSFR-metallicity anti-correlation or a feature of the model converging to an answer at sufficient resolution.
However, it is reassuring that the extent to which there is a mass dependence varies from model-to-model, suggesting that it is not fully explained by numerical convergence (or lack thereof).

At stellar masses greater than than $\sim10^{10.5}~\mathrm{M}_\odot$, we observe an ``inversion'' of $\eta_{\rm SFR}$ for TNG and EAGLE values, where $\eta_{\rm SFR}$ takes on a positive value for the highest mass bins, while the SDSS value increases from negative to consistent with zero~(Figure~\ref{fig:eta0.5}).
This suggests that, in these mass bins, a galaxy's offset from the MZR is {\it positively} correlated with its star formation rate: the opposite of a canonical FMR-like relation. 
In the SDSS data, the inversion is subtle, but in EAGLE and TNG we find strong positive correlations between sSFR and metallicity.
The $\eta_{\rm SFR}$ values for Illustris and SIMBA, on the other hand, remain negative in this mass range.

\section{Discussion} \label{sec:discussion}

In Section \ref{sec:results}, we showed that there exists a relation between stellar mass, gas-phase metallicity, and SFR in the data sources we used for our analysis and that this relation shows a stellar mass dependence, which can evolve with stellar mass.
In this section, we (i) show that this inversion becomes more prominent across our data sources as we include galaxies with lower sSFRs~(Section~\ref{eta2}), (ii) consider whether the choice of SDSS metallicity diagnostic has an impact on our results~(Section~\ref{Z_diagnostic}), (iii) expand our analysis from redshift $z = 0$ to include redshifts up to $z = 1$~(Section~\ref{redshifts}), and (iv) speculate on the physical mechanism(s) that cause this inversion to appear~(Section~\ref{physics}).

\subsection{Changing the Galaxy Selection Criteria Threshold} \label{eta2}

To investigate the impact of low sSFR galaxies on the value of $\eta_{\rm SFR}$, we change the threshold value for the selection criteria (described in Section \ref{selection}). 
In the previous sections, we select galaxies whose sSFR lies above a threshold defined as some fixed offset below the median sSFMS.
The fiducial value of this threshold, $\Delta {\rm sSFR}$, is $-0.5~{\rm dex}$.
In this section, we consider three variations: (i) $\Delta {\rm sSFR}=-0.1~{\rm dex}$, (ii) $\Delta {\rm sSFR}=-1~{\rm dex}$, and (iii) an effective $\Delta {\rm sSFR}=-\infty~{\rm dex}$ cut where we consider all systems with SFR $>0~{\mathrm{M}_\odot\,{\rm yr}^{-1}}$.
Reducing the absolute value of this parameter (i.e., changing it to $-0.1~{\rm dex}$) will include fewer galaxies, as the threshold is more strict. 
Conversely, increasing the absolute value of the parameter (changing it to $-1~{\rm dex}$ or $-\infty~{\rm dex}$) will include more galaxies, including those with lower sSFRs.
This thresholding is done to avoid picking an absolute normalization in sSFR, which can vary from simulation to simulation and the SDSS observations~(Figure~\ref{fig:cut})

\begin{figure*}
    \centering
    \includegraphics[width=\linewidth]{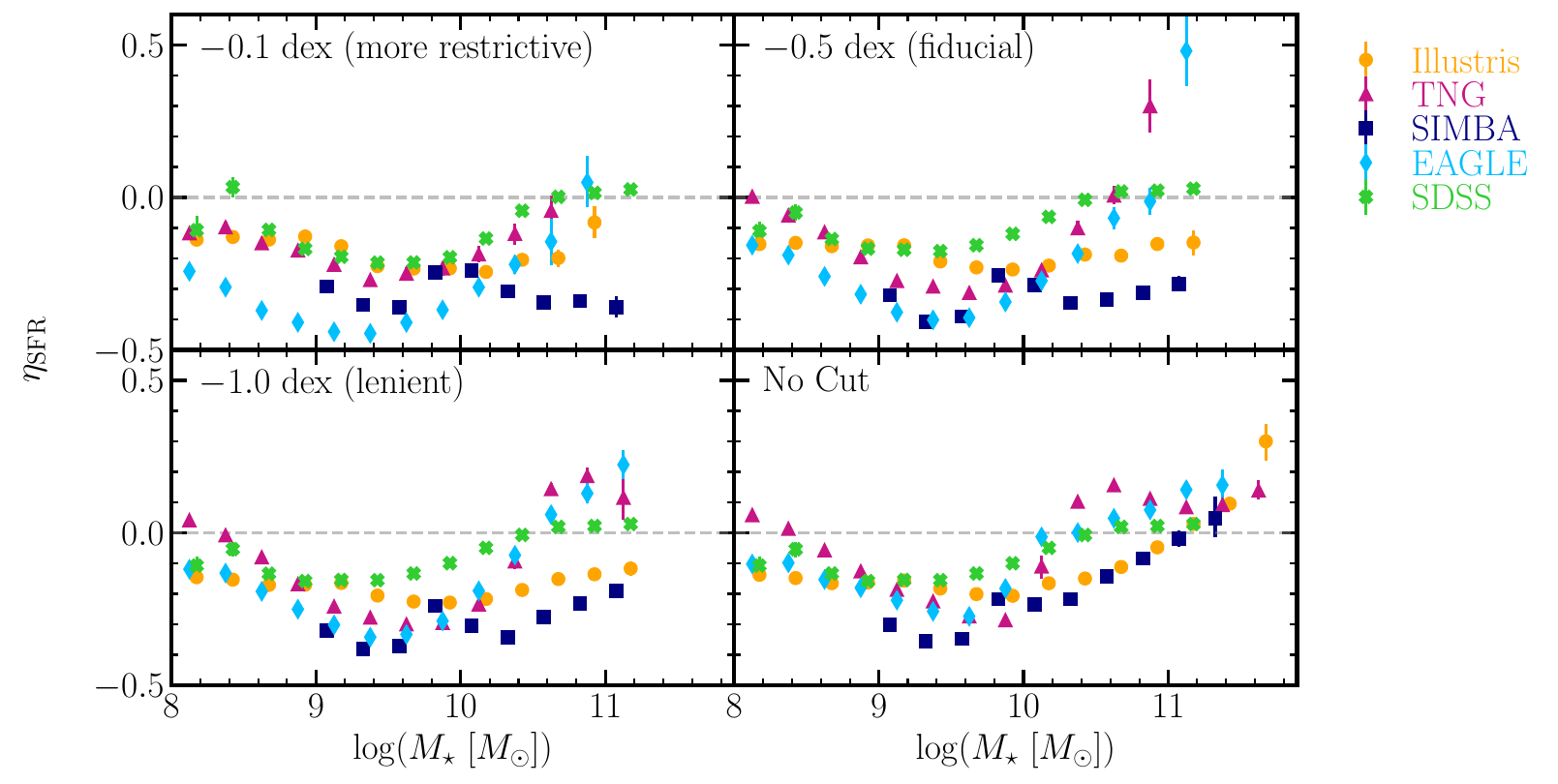}
    \caption{
    {\bf Impact of Low sSFR Galaxies on samples.}
    Comparison of $\eta_{\rm SFR}$ values as a function of mass for mass bins of width $0.25~{\rm dex}$ for each of the four simulations (Illustris, TNG, SIMBA, EAGLE) and SDSS with four different sSFR selection threshold values: selecting all galaxies greater than $-0.1$ dex above the sSFMS (top left), greater than $-0.5$ dex (fiducial, top right), greater than $-1$ dex (bottom left), and making no cut on the sSFRs (bottom right).
    \label{fig:etas}
    }
\end{figure*}

Figure~\ref{fig:etas} shows the equivalent of Figure \ref{fig:eta0.5} for each of the three changes in threshold value.
The top left panel depicts the value of $\eta_{\rm SFR}$ as a function of mass for a threshold of $\Delta {\rm sSFR} = -0.1~{\rm dex}$.
With this highly restrictive sSFR cut, for virtually every mass bin and every data source, we find that $\eta_{\rm SFR}$ is negative.
Interestingly, there are a few outliers (one low mass bin for SDSS and the highest mass bins for SDSS and EAGLE), but even these are only slightly greater than zero and the EAGLE mass bin's uncertainty is large.
The top right panel is identical to Figure \ref{fig:eta0.5}, we show this panel again here as a point of reference~(see Section \ref{eta1} for discussion of Figure~\ref{fig:eta0.5}).

Both bottom panels depict $\eta_{\rm SFR}$ as a function of mass bin for relaxed selection criteria ($\Delta {\rm sSFR}=-1.0~{\rm dex}$ in the bottom left and $\Delta {\rm sSFR}=-\infty,~{\rm i.e.,}~{\rm SFR}>0$, in the bottom right).
In both panels, we observe a similar pattern as in the top right panel (Figure \ref{fig:eta0.5}) for SDSS, TNG, and EAGLE:
for stellar masses between $10^8$ and $10^{10.5}~\mathrm{M}_\odot$, the $\eta_{\rm SFR}$ values are negative and approach zero as the mass increases;
for $M_\star\gtrsim10^{10.5}~\mathrm{M}_\odot$ mass bins, EAGLE and TNG show positive $\eta_{\rm SFR}$ values, while SDSS  $\eta_{\rm SFR}$ values are consistent with $\sim0$.
The behavior of Illustris and SIMBA for the slightly less restrictive case ($\Delta {\rm sSFR}=-1.0~{\rm dex}$) is the same as for the fiducial threshold, with negative $\eta_{\rm SFR}$ values at every mass.
Remarkably, however, this changes when considering all ${\rm SFR}>0$: at $M_\star \gtrsim 10^{11}\mathrm{M}_\odot$, both Illustris and SIMBA have positive $\eta_{\rm SFR}$ values~(although the uncertainty on the positive SIMBA $\eta_{\rm SFR}$ includes zero). 
Therefore, using this least restrictive $\rm \Delta {\rm sSFR} = -\infty$ cut, we find a flattening/inversion to some extent of $\eta_{\rm SFR}$ at the highest masses for all five data sources.

We suggest that these findings are significant in the following way:
when including only high sSFR galaxies ($\Delta {\rm sSFR} = -0.1~{\rm dex}$), $\eta_{\rm SFR}$ values are negative and we do not see an inversion at higher masses~(with the exception of a few weak, low-significance outliers).
Interestingly, we start to see the inversion of $\eta_{\rm SFR}$ at higher masses for some of our data sources when we relax the galaxy selection threshold to include some lower sSFR galaxies ($\Delta {\rm sSFR} = -0.5, -1.0 ~{\rm dex}$).
Once we relax this threshold all the way to include all galaxies that have a non-zero sSFR ($\Delta {\rm sSFR} = -\infty ~{\rm dex}$), we see the inversion of $\eta_{\rm SFR}$ at higher masses for all four of the simulations.

We note that the SDSS values are less affected by the variations of the threshold as there are fewer low sSFR galaxies in the data set (we restrict our sample to galaxies with high S/N emission lines).
In the case of SDSS, then, lowering the threshold to include more low sSFR galaxies has little effect because the cuts disqualify fewer galaxies to begin with.
The agreement between the four simulations when lowering the threshold suggests that this inversion is likely related to the presence of these low sSFR galaxies.

\subsection{Choice of Metallicity Diagnostic with SDSS Data} \label{Z_diagnostic}

\begin{figure}
    \centering
    \includegraphics[width=\linewidth]{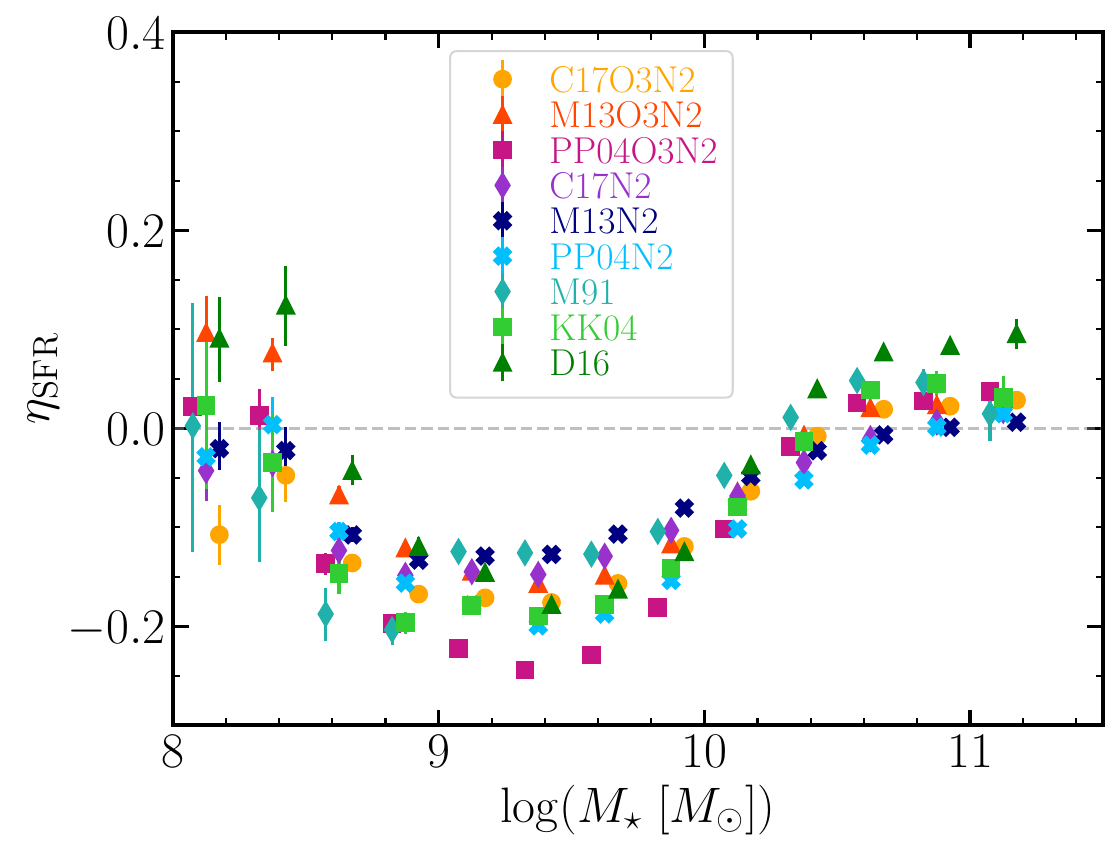}
    \caption{{\bf Impact of Metallicity Diagnostic.}
    Comparison of $\eta_{\rm SFR}$ values as a function of mass for mass bins of width $0.25~{\rm dex}$~(using the fiducial sSFMS selection criteria) at $z = 0$ for $9$ different metallicity diagnostics.
    We note that the C17O3N2 diagnostic is the fiducial diagnostic we use in this work.
    See also Appendix~\ref{appendix:diagnostics} and Figure~\ref{fig:diagnostics_MZR} full MZRs from each diagnostic and comparisons to simulations.
    }
    \label{SDSS}
\end{figure}

While the simulations we use allow us to directly obtain specific values for metallicity, this is not as easily done when calculating metallicity from observations.
In our analysis, we use the $12+\log({\rm O/H})$ values from the $\rm O_3N_2$ metallicity diagnostic as per \cite{Curti_2017}.
However, as shown in \citeauthor{Andrews_2013} (\citeyear{Andrews_2013}; see their Table 5) the residual correlation about the MZR can potentially depend on the adopted metallicity diagnostic.
Therefore, in an effort to quantify how much the chosen metallicity diagnostic plays a role in our observational results, we recreate our analysis using nine different metallicity diagnostics.
Specifically, the different metallicity diagnostics we use are as follows: $\rm O_3N_2$ and $\rm N_2$ indicators from \citeauthor{Curti_2017} (\citeyear{Curti_2017}; hereafter
\textit{C17O3N2} and \textit{C17N2}, respectively), \citeauthor{Marino_2013} (\citeyear{Marino_2013}; \textit{M13O3N2} and \textit{M13N2}, respectively), and \citeauthor{Pettini_2004} (\citeyear{Pettini_2004}; \textit{PP04O3N2} and \textit{PP04N2}, respectively), as well as the \citeauthor{McGaugh_1991} (\citeyear{McGaugh_1991}; \textit{M91}), \citeauthor{Kobulnicky_2004} (\citeyear{Kobulnicky_2004}; \textit{KK04}), and \citeauthor{Dopita_2016} (\citeyear{Dopita_2016}; \textit{D16}) diagnostics.
In this section, we compare the correlations of the offsets with respect to sSFR; however, we present the full MZRs from each diagnostic in Appendix~\ref{appendix:diagnostics}~(Figure~\ref{fig:diagnostics_MZR}).

Each diagnostic generally uses strong-line-methods, which entail calibrating the ratio of lines with a known metallicity dependence against the oxygen content of the galaxy.
\cite{Curti_2017} combine the strong-line-method with the use of electron temperature (anti-correlated with metallicity) and stacking galaxy spectra to empirically obtain metallicity values.
Similarly, \cite{Marino_2013} combine the strong-line-method and the use of electron temperature in their calibration.
\cite{Pettini_2004} also use the strong-line-method.
In the cases of \cite{McGaugh_1991}, \cite{Kobulnicky_2004}, and \cite{Dopita_2016}, this calibration was done in a more theoretical way using photoionization models.

We show the comparison of the $\eta_{\rm SFR}$ versus stellar mass in Figure~\ref{SDSS}~(analogous to Figure~\ref{fig:eta0.5}) for each of the nine diagnostics.
We note that for this analysis we use the fiducial galaxy selection threshold ($\Delta {\rm sSFR} = - 0.5~{\rm dex}$).
We generally find strong agreement between each of the different diagnostics.
The $\rm \eta_{SFR}$ values follow the same trend: positive for the lowest mass bin(s), negative for $10^{8.5} \mathrm{M}_\odot < M < 10^{10.5} \mathrm{M}_\odot$, and then inverting to positive at masses greater than $10^{10.5} \mathrm{M}_\odot$.\ignorespaces
\footnote{\ignorespaces
We note that for the M91 and KK04 diagnostics, some of the increased uncertainty in $\eta_{\rm SFR}$ at low masses ($M_\star\lesssim 10^{8.5}~{\rm M}_\odot$) comes from apparent ``clouds'' of low metallicity systems~(see bottom left and bottom middle panel of Figure~\ref{fig:diagnostics_MZR}).
}
In more detail, the extent to which an inversion occurs at the highest masses does seem to appear.
For example, the D16 diagnostic seems to present the strongest positive correlations, whereas many of the other diagnostics~(especially PP04N2, M91, and M13N2) are more consistent with zero.
Moreover, the stellar mass at which $\eta_{\rm SFR}$ inverts can be different from diagnostic-to-diagnostic.
For the M91 and D16 diagnostics, the inversion first occurs at $\sim10^{10.4}~{\rm M}_\odot$, and the inversion seems to have substantially less significance at the highest masses~($M_\star\sim10^{11}~{\rm M}_\odot$) for M91.
For KK04, M13O3N2, PP04O3N2, and C17O3N2, the inversion happens at $\sim10^{10.6}~{\rm M}_\odot$.
Finally, C17N2, M13N2, and PP04N2 have only a weak inversion~(virtually consistent with zero) at $\sim10^{10.8}~{\rm M}_\odot$.
It is reassuring that the different tracers, e.g., $\rm O_3N_2$ and $\rm N_2$, show similar trends.
We therefore conclude that while the details of exactly when, where, and how much the inversion occurs may depend on chosen diagnostic, the qualitative finding of the existence of some mass evolution with a flattening/inversion  in $\eta_{\rm SFR}$ for high mass systems seems to be a generic prediction of the observed data.

\subsection{Persistence of Inversion to $z=1$ in Simulations} \label{redshifts}

\begin{figure*}
    \centering
    \includegraphics[width=0.85\linewidth]{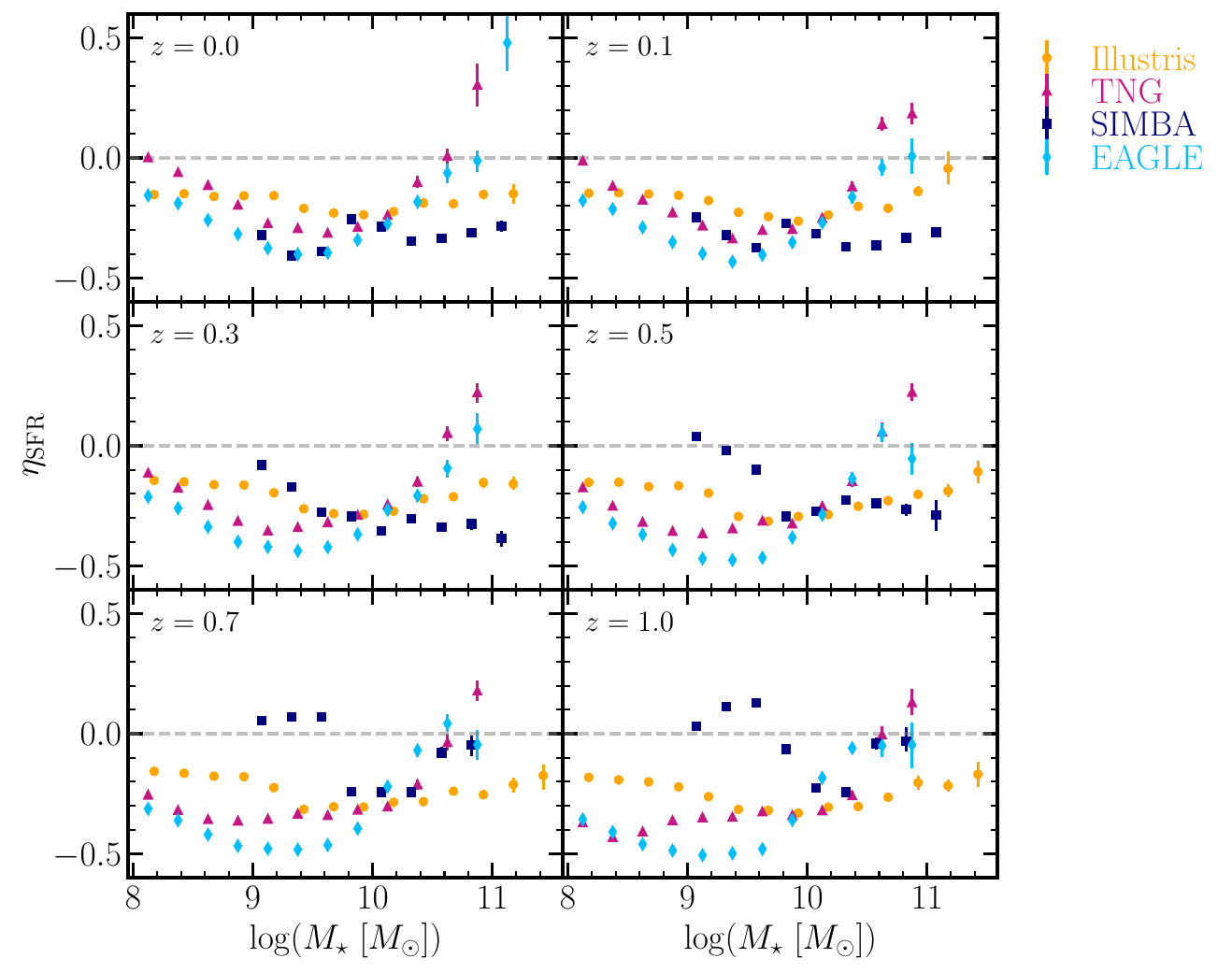}
    \caption{
    {\bf Redshift Evolution of $\eta_{\rm SFR}$ in Cosmological Simulations.}
    Comparison of $\eta_{\rm SFR}$ as a function of mass for mass bins of width $0.25~{\rm dex}$ using $\Delta {\rm sSFR} = -0.5~{\rm dex}$ for each of the four simulations (Illustris, TNG, SIMBA, EAGLE) at 6 different redshifts: $z = 0$ (top left, equivalent to Fig. \ref{fig:eta0.5}), $z = 0.1$ (top right), $z = 0.3$ (middle left), $z = 0.5$ (middle right), $z = 0.7$ (bottom left), and $z = 1$ (bottom right). 
    }
    \label{redshift}
\end{figure*}

In addition to $z=0$, we perform the same analysis described in Sections \ref{eta1} and \ref{eta2}, using simulation data in different increments up to redshift $z = 1$.
Figure \ref{redshift} shows $\eta_{\rm SFR}$ values for the four simulations we use at redshifts $z = 0$, $0.1$, $0.3$, $0.5$, $0.7$, and $1.0$.
We note that this analysis was done using the fiducial threshold of $\Delta {\rm sSFR} = -0.5~{\rm dex}$; however, we find that the pattern of less restrictive sSFR criteria leading to a stronger high-mass inversion established in Section \ref{eta2} also hold for all redshifts up to $z = 1$ in the simulations.
As a point of reference, we also include the SDSS data in the $z=0$ panel.

For TNG and EAGLE, the behavior established at $z = 0$ (see Section \ref{eta1}) persists at higher redshifts:
negative $\eta_{\rm SFR}$ values for mass bins up to $\sim10^{10.5} \mathrm{M}_\odot$ with an inversion to positive $\eta_{\rm SFR}$ in the highest mass bins.
Moreover, the relatively constant $\eta_{\rm SFR}$ values found in Illustris at $z=0$ continue at $z>0$; however, in a few of the highest mass bins at $z>0$ (e.g., $M_\star\sim10^{11}~{\rm M}_\odot$ at $z=0.1$), there are hints of deviations from this constant $\eta_{\rm SFR}$.
SIMBA seems to follow similar behavior at $z=0.1$ and $0.3$ to that observed at $z=0$.
At $z\geq 0.5$, there appears an inversion of $\eta_{\rm SFR}$ on the low mass, with positive $\eta_{\rm SFR}$ values at masses below $10^{10}~\mathrm{M}_\odot$ and negative $\eta_{\rm SFR}$ values above $10^{10}~\mathrm{M}_\odot$, the origin of which is unclear but is unique to SIMBA.
The fiducial SIMBA model does have some redshift scaling winds, however, these are only scaled at $z>3$.
Indeed, it seems that the $\eta_{\rm SFR}$ values generally become slightly less negative than their lower redshift counterparts across virtually all mass bins.
While an interesting quirk of the SIMBA model, we leave a more thorough investigation of the detailed feedback histories of SIMBA galaxies to another work and simply note here that there is some evolution of $\eta_{\rm SFR}$ in SIMBA, moreso than the other three simulations.

\subsection{Possible Origin of High-Mass Inversion} \label{physics}

\begin{figure*}
    \centering
    \includegraphics[width=\linewidth]{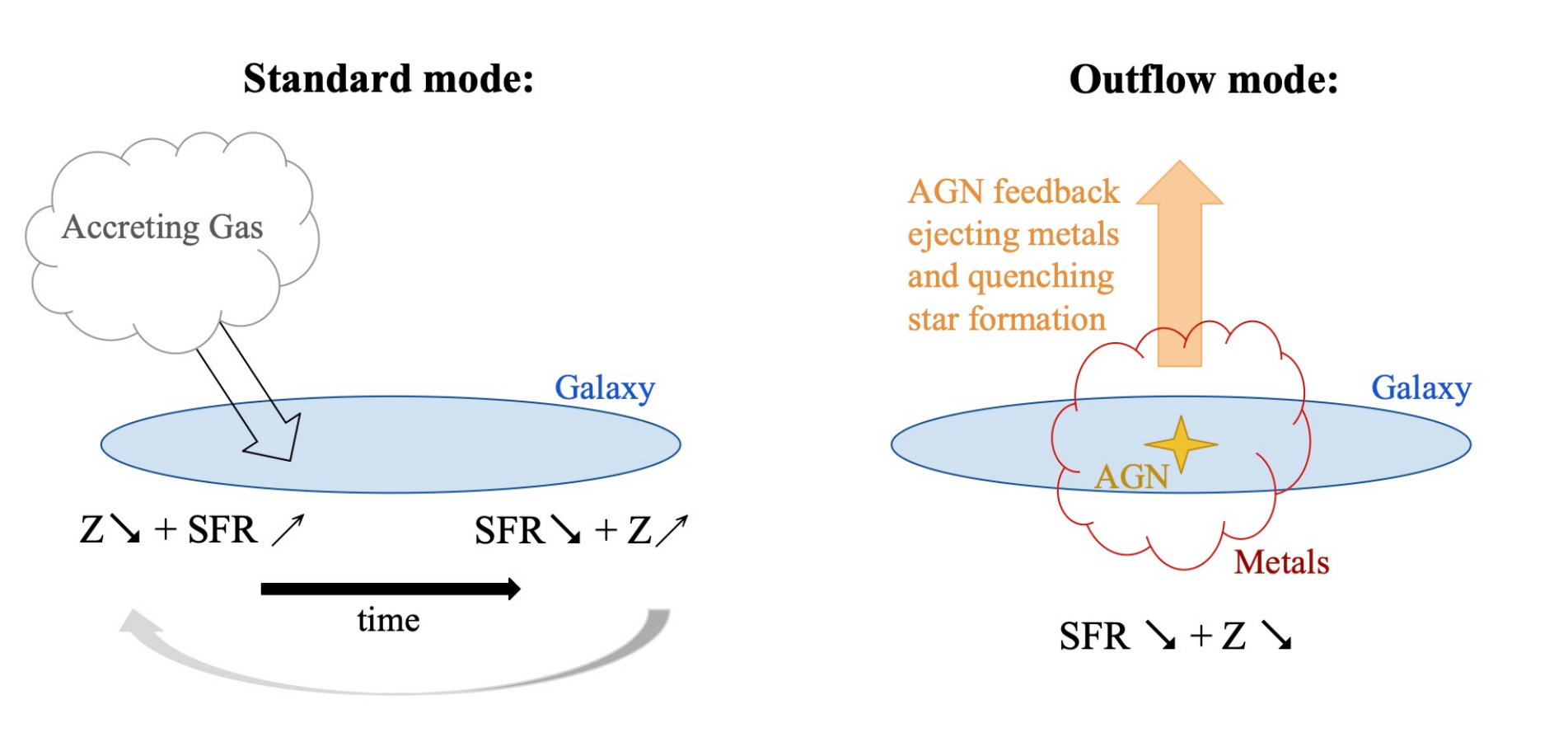}
    \caption{
    {\bf Cartoon Picture of Possible Origin of $\eta_{\rm SFR}$ Inversion.}
    Simplified models of the physical processes involved in the `standard' and proposed AGN modes. \textit{Left:} `Standard' FMR model, wherein as gas is accreted, the metallicity decreases and SFR increases. Once the gas is used up and stars die and release metals, the SFR decreases and the metallicity increases. 
    \textit{Right:} Proposed Outflow/AGN mode, showing a higher concentration of metals near the center of a galaxy, where they are preferentially blown out by (likely) AGN feedback in the simulations, resulting in metallicity and SFR decreasing at the same time.
    }
    \label{graphic}
\end{figure*}

A common physical picture suggested in the ``standard'' version of the anti-correlation between metallicity and SFR follows a simple gas regulator-type model~\citep[e.g.,][]{Lilly_2013}.
A galaxy accretes pristine gas, lowering its metallicity, which then fuels the birth of new stars, increasing its SFR.
As this gas is being used up, the SFR decreases, and as the stars die, they release metals into the galactic environment, raising the metallicity.
The result of this process is a negative $\eta_{\rm SFR}$ value, like those in the insets of Figure \ref{fig:MZR}.
We present a simple schematic for this ``standard mode'' is depicted in the left panel of Figure \ref{graphic}.

However, the simplistic picture of the ``standard mode'' alone does not explain the inversion to positive $\eta_{\rm SFR}$ values at higher mass bins.
We suggest that the inversion (i.e., positive $\eta_{\rm SFR}$ values) noted in the simulations and observations might be the result of systems experiencing a strong nuclear outflow event driving metal rich gas from the galaxy that is simultaneously preventing new stars from forming.
This ``outflow mode'' would leads to a positive correlation between the two quantities, i.e., a positive $\eta_{\rm SFR}$ value: as the metallicity is driven down by the outflowing gas, so too is the star formation rate.
Thus we speculate that this inversion of the anti-correlation is one-directional, such that raising the metallicity does not necessarily correlate with a highly star forming system.

One such mechanism by which these powerful outflows can be fueled is with AGN.
In the simulations, galaxies with larger masses have some prescription for AGN feedback, which produce outflows \citep{McAlpine_2017,Weinberger_2017,Torrey_2019,Wright_2024}.
Specifically, \cite{Torrey_2019}, looking at how galaxies in the TNG simulation move from redshift $z=0.058$ to $z=0$, have shown that at masses $> 10^{10.5}~\mathrm{M}_\odot$, the galaxies go from being accretion- (increasing ISM mass, decreasing metallicity) or enrichment-dominated (decreasing ISM mass, increasing metallicity) to being outflow-dominated (decreasing ISM mass, constant metallicity; see Figure 13 in \citeauthor{Torrey_2019} \citeyear{Torrey_2019}).
Higher mass galaxies in this case lose ISM mass while maintaining a constant metallicity, meaning they entrain both hydrogen and metals within the outflows.
Similarly, \cite{Wright_2024} show that in TNG, EAGLE, and SIMBA, AGN-driven outflows are expelled to approximately $R_{\rm200c}$ (the radius at which the density becomes $200\times \rho_{\rm critical}$) or further.
However, in these simulations, quantities like gas-phase metallicity are measured over a radius of around $2 R_\star$, which is significantly smaller than $R_{\rm200c}$.
The AGN, being located in the center of their galaxy, will likely preferentially eject material located in the center of the galaxy, which tends to be more metal-enriched than the outskirts \citep[e.g., they have negative metallicity gradients][]{Garcia_2025b,Garcia_2026}.
Indeed, there is some evidence from spatially resolved integral field unit~(IFU) observations that AGN outflows can entrain metal-rich gas~\citep[][Zhu et al. Submitted, Kr{\'o}l et al. In Preparation]{Kirkpatrick_2009,Villar-Martin_2024,Krol_2026}.
However, there is not a clear, ubiquitous picture of the role of AGN feedback in galactic centers in observed galaxies~\citep{do-Nascimento_2022,Amiri_2024,Amiri_2025}.
The extent to which an inversion of $\eta_{\rm SFR}$ persists at high masses is therefore a potentially interesting test of the simulations' AGN feedback prescriptions.
Too strong of AGN feedback might preferentially blow-out metals from the galactic center leading to stronger population-level $\eta_{\rm SFR}$ inversions, whereas too weak feedback might prohibit the flattening/inversion at higher masses.

\cite{DeRossi_2017} also used the EAGLE simulation to investigate the effect of AGN on the MZR and FMR.
They found that the AGN feedback starts to have an influence at masses $> 10^{10} \mathrm{M}_\odot$, around the same threshold point where the inversion of $\eta_{\rm SFR}$ in our results occurs.
In the lowest panel of their Figure 3, they show that if no AGN is present, the metallicity continues to increase with stellar mass. 
However, if an AGN is present, the metallicity plateaus or decreases at masses $> 10^{10} \mathrm{M}_\odot$, as metals are ejected from the galactic environment.
\citeauthor{DeRossi_2017} show that the presence of AGN (feedback) has an influence on the behavior of the MZR/FMR via the AGN's effect on the metallicity evolution.
This, in effect, is what we are suggesting is occurring here in the simulated datasets~(see also above discussion on observed galaxies).
The inversion of $\eta_{\rm SFR}$ values occurs at a mass of $10^{10.5}~\mathrm{M}_\odot$ --- roughly the threshold where galaxies begin to be outflow-dominated and AGN feedback has more of an influence \citep{Torrey_2019, DeRossi_2017}.
We propose that at higher masses, the presence of AGN causes outflows of mainly metal-enriched material, as that is more concentrated in the center of the galaxy \citep{Garcia_2025b}, which are ejected far enough to no longer be counted when calculation the galaxy's metallicity \citep{Wright_2024}.
At the same time, the presence of the AGN causes a decrease in sSFR~(by construction in the simulations).
The simultaneous decrease in metallicity and sSFR at masses $\gtrsim 10^{10.5} \mathrm{M}_\odot$ then leads to positive $\eta_{\rm SFR}$ values.
A simplified depiction of this mechanism (`outflow mode') can be found on the right panel of Figure~\ref{graphic}.
Our analysis in Section~\ref{eta2} and Figure~\ref{fig:etas} lends further support to this suggestion (at least, for the simulations):
increasing the sSFR of the galaxies we select leads to a decrease in inversion, while selecting galaxies with lower sSFRs increases the presence of the inversion, tying the presence/strength of the inversion to the sSFR of the galaxies, potentially further linking the inversion to outflows powered by strong nuclear stellar/AGN feedback.

Briefly, we also saw positive correlations between sSFR and metallicity in low mass systems in TNG and SIMBA.
This picture is qualitatively similar to that of the AGN case, though---at least in the simulations---requires a different physical explanation.
We speculate that these were driven by increased efficacy of the stellar feedback in these regimes within the two models~\citep[see][as well as discussion in Section~\ref{eta1}]{Pillepich_2018,Davé_2019}.
The shallower gravitational potentials make it more likely for these low mass system to lose their gas; however, it is not clear exactly why this would lead to preferential metal rich gas outflows.
Regardless, the strong stellar feedback at these masses (and redshifts) in these models is a possible scenario in which a simultaneous decrease in metallicity and SFR can occur in low-mass galaxies.

\section{Conclusions} \label{sec:conc}

In this paper, we investigate the mass dependence of the residual correlation with sSFR about the MZR using data from the cosmological simulations EAGLE, SIMBA, Illustris, and TNG, as well as from SDSS observational data.
We restrict our samples to star forming galaxies with masses ranging from $10^{8}~{\rm M}_\odot < M_\star < 10^{12}~{\rm M}_\odot$.
We also expand our analysis from redshift $z = 0$ in increments up to $z = 1$.
We reach the following conclusions:

\begin{itemize}

    \item We find that there is generally an anti-correlation between offsets from the MZR based on sSFR at $z = 0$, with lower metallicities corresponding to higher sSFRs, and vice-versa (Figure \ref{fig:MZR}).
    We quantify this anti-correlation for the entirety of each sample by defining $\eta_{\rm SFR}$ as the slope of the relation between $\Delta Z$ and sSFR (Equation \ref{eq1}).
    We find that the $\eta_{\rm SFR}$ is negative when considering all galaxies in each sample (e.g., insets of Figure \ref{fig:MZR}).
    
    \item We next perform a similar experiment breaking down the samples into thin mass bins of width $0.25~{\rm dex}$.
    We find that $\eta_{\rm SFR}$ for these individual mass bins is generally negative; however, there is an inversion of $\eta_{\rm SFR}$ at higher masses ($\rm M_\star \gtrsim 10^{10.5} \mathrm{M}_\odot$) in SDSS, TNG, and EAGLE~(Figure \ref{fig:eta0.5}).

    \item We show that adjusting our galaxy selection criteria by changing the SFR threshold from its fiducial value of $\Delta {\rm sSFR} = -0.5~{\rm dex}$ below the sSFR main sequence has an effect on the inversion of $\eta_{\rm SFR}$ (Figure \ref{fig:etas}). 
    Restricting our sample to include fewer low-sSFR galaxies ($\Delta {\rm sSFR} = -0.1~{\rm dex}$) reduces the inversion, whereas relaxing the threshold to include more low-sSFR galaxies ($\Delta {\rm sSFR} = -1, -\infty~{\rm dex}$) makes the inversion more pronounced. Using $\Delta {\rm sSFR} = -\infty~{\rm dex}$, the inversion of $\eta_{\rm SFR}$ is present for all five data sources at the highest masses. 
    We conclude that the inversion is related to, and caused by, the presence of low sSFR galaxies.

    \item We find that our results are independent of the choice of SDSS metallicity diagnostic (Figure \ref{SDSS}), suggesting that our observational results are robust to uncertainties in metallicity calibration in observations.

    \item We find that the established trend (inversion of $\eta_{\rm SFR}$ values at higher masses for EAGLE and TNG, negative $\eta_{\rm SFR}$ values for Illustris) remains present at redshifts up to $z = 1$ (Figure \ref{redshift}). We note that the behavior of SIMBA $\eta_{\rm SFR}$ values changes at $z \gtrsim 0.5$, although a more detailed analysis of this phenomenon lies outside the scope of this paper.

    \item We propose an alternative physical picture to the `standard' model used to explain the anti-correlation of metallicity and sSFR, which would account for the resulting positive correlation between SFR and metallicity of high-mass galaxies (Figure \ref{graphic}).
    Specifically, we suggest that strong nuclear outflows (from, e.g., AGN feedback) simultaneously reduce star formation (lowering the SFR) and preferentially blow out gas from the metal rich centers of galaxies.
    
\end{itemize}

We have shown that there exists a residual correlation about the MZR that changes its behavior at higher masses ($\rm M \gtrsim 10^{10.5}~\mathrm{M}_\odot$) across simulations (Illustris, TNG, EAGLE, SIMBA) up to at least $z = 1$ and local universe observations with SDSS.
The ubiquity of the \mbox{(anti-)correlations} of metallicity and SFR in our models and observations gives hints towards an larger picture of galactic scale physics.
Yet, treating these subtle processes with a single mass and redshift invariant relationship glosses over critical pieces of physics.
Having a clearly quantified way to measure the anti-correlation between metallicity and sSFR as a function of mass allows for a more robust understanding of the story metals and SFR tell us.
This framework allows us to more fully see the connection of the inversion to the inclusion of low sSFR galaxies, which in turn allows for a new physical description of this relation in high-mass galaxies.
It also provides a means of understanding mechanisms that quench galaxies.
This can put future constraints on implementing structures that influence the behavior of metallicity in simulations, like AGN feedback or mass loading factors, which would not be possible by treating the FMR in a rigid framework.

\begin{acknowledgements}
The authors acknowledge Research Computing at The University of Virginia for providing computational resources and technical support that have contributed to the results reported within this publication. URL: \href{https://rc.virginia.edu}{https://rc.virginia.edu}.

AMG and PT acknowledge support from NSF-AST 2346977 and the NSF-Simons AI Institute for Cosmic Origins which is supported by the National Science Foundation under Cooperative Agreement 2421782 and the Simons Foundation award MPS-AI-00010515.

KG is supported by the Australian Research Council through the Discovery Early Career Researcher Award (DECRA) Fellowship (project number DE220100766) funded by the Australian Government. 

JMS acknowledges support from the National Science Foundation under Grant No 2205551.

\end{acknowledgements}

\appendix

\section{MZR from different Diagnostics}
\label{appendix:diagnostics}

\setcounter{equation}{0}
\setcounter{figure}{0} 
\setcounter{table}{0}
\renewcommand{\theequation}{A\arabic{equation}}
\renewcommand{\thefigure}{A\arabic{figure}}
\renewcommand{\thetable}{A\arabic{table}}

\begin{figure*}
    \centering
    \includegraphics[width=\linewidth]{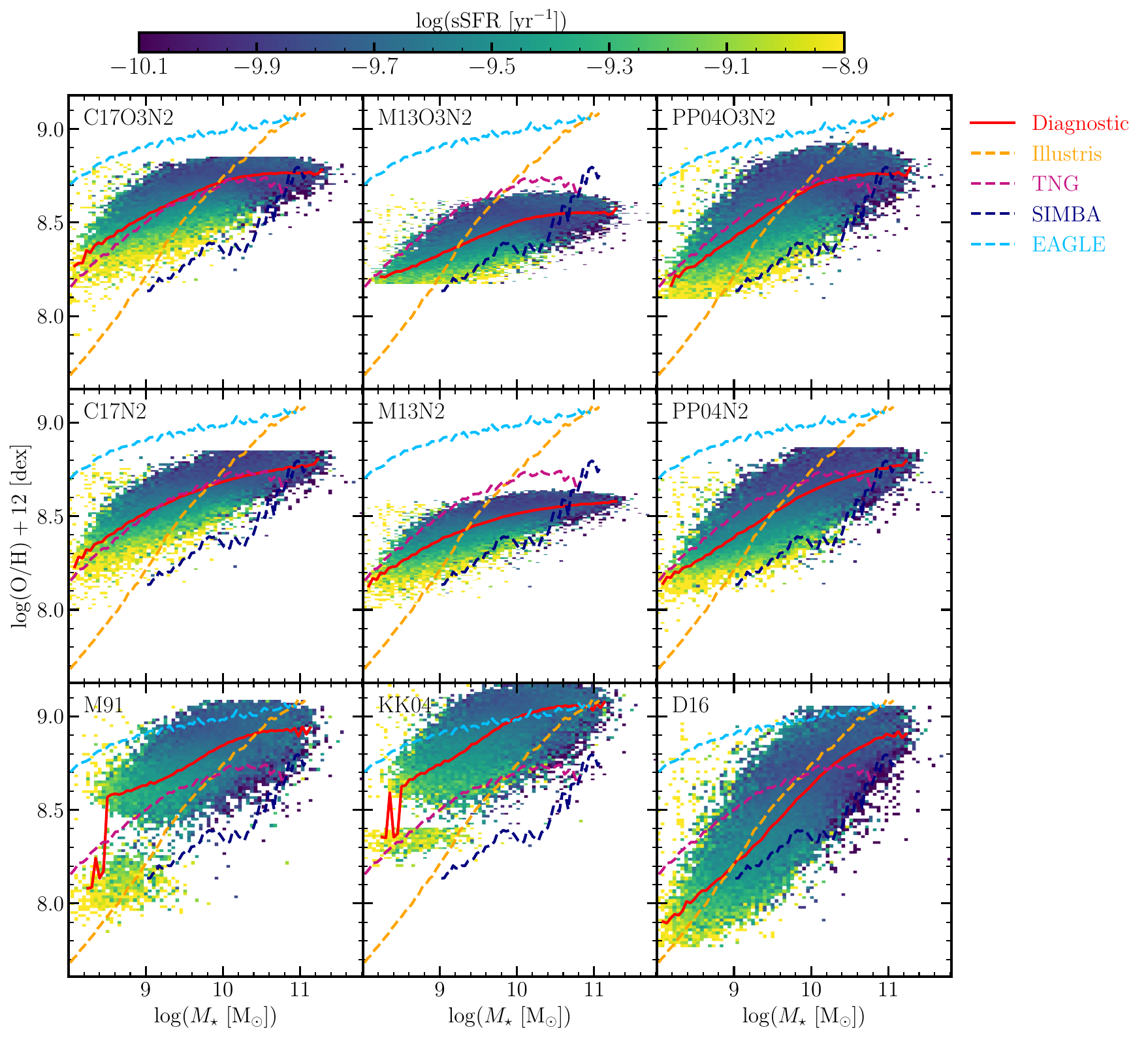}
    \caption{{\bf Comparison of Different MZRs from Each Metallicity Diagnostic.}
    The background histograms show the distribution of galaxies from our SDSS sample with different metallicity diagnostics (as labeled in the upper left-hand portion of each panel).
    The solid red line is a rolling median for each diagnostic.
    The dashed lines are those from the simulations~(see also Figure~\ref{fig:MZR}).
    }
    \label{fig:diagnostics_MZR}
\end{figure*}

In this appendix, we present the full MZRs from each of the nine different metallicity diagnostics in our SDSS sample from Section~\ref{Z_diagnostic}.
Figure~\ref{fig:diagnostics_MZR} shows a 2D histogram of the full distribution of galaxies from the SDSS sample in measured with of the nine diagnostics.
Similar to Figure~\ref{fig:MZR}, the bins in each panel are color-coded by the average sSFR of galaxies in each bin.
The solid red line in each panel is a rolling median of the metallicity in different stellar mass bins.
We also overplot the rolling medians from the four simulations~(same data as in Figure~\ref{fig:MZR}).
The abrupt upper edges in the panels emerge from the samples reaching the maximum viable metallicity for their respective metallicity diagnostic calibration.

As noted a number of times previously in the metallicity diagnostic literature~(most famously in~\citeauthor{Kewley_2008}~\citeyear{Kewley_2008}), we find that both the absolute normalization and shape of the MZR vary from one diagnostic to another.
However, we note that in each of the panels, a visual inspection of the correlation with sSFR yields the expected anti-correlation between MZR offsets and sSFR~(see Section~\ref{Z_diagnostic} for more quantitative exploration).
We further emphasize that the result of TNG having the closest normalization and shape of the MZR to the SDSS sample~(Section~\ref{fmr}) is subject to the specific diagnostic chosen.
As an example, if we were instead to use the D16 diagnostic~(bottom right panel of Figure~\ref{fig:diagnostics_MZR}), both the shape and normalization of the Illustris MZR match with the SDSS sample best.
It is therefore difficult to make concrete statements about the (dis-)agreement between the observed sample and simulations.
This further highlights the utility of comparing their scatters, which seem to be much less sensitive to the chosen diagnostic~(Figure~\ref{SDSS}).

\bibliography{sample631}{}
\bibliographystyle{aasjournal}

\end{document}